\documentclass[prb,showkeys,superscriptaddress,byrevtex,a4paper,twocolumn,10pt]{revtex4}
\usepackage{dcolumn}
\usepackage{bm}
\usepackage[latin2]{inputenc}
\usepackage{graphics}
\usepackage{longtable}
\usepackage{amsmath}
\usepackage{marvosym}
\usepackage{pifont}
\usepackage{bbm}
\usepackage{amssymb}

\newcommand{\BE}{\begin{equation}}
\newcommand{\EE}{\end{equation}}
\newcommand{\BEA}{\begin{eqnarray}}
\newcommand{\EEA}{\end{eqnarray}}

\newcommand{\BM}[1]{\mbox{\boldmath$#1$}}
\begin{document}
\title{Solution of the Poisson equation for two dimensional 
  periodic structures (slabs) in an overlapping localized site density scheme}

\author{F. Tasn\'adi}
\email[E-mail: ]{f.tasnadi@ifw-dresden.de}
\affiliation{IFW
   Dresden e.V.,\\  P.O. Box 270 116, D-01171 Dresden, Germany}

\date{\today}

\begin{abstract}
Bertaut's equivalent electric density idea (E. F. Bertaut,
Journal de Physique {\bf 39}, 1331 (1978)) is applied 
to the case of two dimensional periodic continuous charge density
distributions. The following derivation differs from what was introduced by Bertaut.
The presented method solves the Poisson equation
for the scheme of overlapping localized site densities with periodic boundary
conditions in the ($x,y$) plane and with the general finite voltage
boundary condition in the perpendicular $z$-direction. 
As usual the long-range potential is calculated in the
Fourier space. For the $K_{||}\ne 0$ case a Fourier transformation helps
to calculate 
the solution in a three dimensional periodic sense, while for 
$K_{||}=0$ 
the required charge neutrality is the starting
point. 
For both
cases suitable representations of the spherical harmonics are needed to
arrive at expressions that are convenient for numerical implementation.
In this localized density scheme an explicit relation can be
derived between the finite voltage in $z$-direction and the $z$-component of the dipole
density.
\end{abstract}

\keywords{slabs, Ewald summation, Poisson equation}
\maketitle 

\section{Introduction}
%
%
The solution of the Poisson equation for periodic structures has great
practical importance in many areas of solid state physics, e.g.\ in band
structure calculations or in molecular dynamics simulations. 
The Bertaut's '{\it equivalent electric density}' idea
\cite{Bertaut}, realized in the pseudo-charge or multipole compensation method
developed by Weinert\cite{Weinert} is nowadays the standard
approach in plane-wave implementations for three dimensional (3D)
periodic problems.
The Bertaut approach to sum up the long-range electrostatic
interactions is based on the replacement of the point-like multipoles at
every lattice sites by non overlapping charge distributions with
equivalent multipoles. In the Weinert method every site might
have several multipoles and the crystal is divided up into atomic
site domains $\Omega$ (muffin-tin sphere) and into an interstitial part.
To calculate the electrostatic potential in the interstitial region (the
long-range potential contribution) Weinert realizes Bertaut's idea by introducing 
pseudo-charge densities confined to the atomic volumes $\Omega$.
Inside $\Omega$ the Dirichlet problem for the original charge density is
solved, using the potential value at the boundary of $\Omega$, which result from
the interstitial solution.
%
%

Dealing with 3D crystalline extended systems the periodic Born-von
K\'arm\'an boundary conditions (PBC) in each direction are mandatory.
For films or single slab geometries \cite{CrystallographyE} the PBC is
applicable only to the extended $(x,y)$ (in-plane) directions. In the
perpendicular $z$-direction the system has only finite extension and
requires a different boundary condition. The most general physical
situation allows for a finite voltage at infinity in $z$-direction,
which would be the case for slabs without a in-plane mirror symmetry.
This general finite voltage condition in the $z$-direction coincides
with the physical picture of a charged capacitor.

%
%

A possible treatment of films is the super-cell (vacuum-film-vacuum)
method with 3D periodicity, which describes the electrostatic potential
inside the super-cell as a continuous function and thus excludes any
step like discontinuities.
%
%
The introducing of an artificial dipole layer
in the vacuum region\cite{Neugebauer,Bengtsson} allows to model the
finite voltage situation. However, there is still a possibility of
spurious  image interactions. Furthermore, an additional parameter, the vacuum
thickness, has to be converged. 
The plane-wave based layer-FLAPW method
\cite{Appelbaum-Hamann,Wimmer} treats the single slab geometry
as a real 2D periodic unit cell extending from $-\infty$ to $+\infty$ in $z$-direction with
three different regions: the muffin-tin, the interstitial, and a vacuum
(no charge) segment. 
The method is applicable to surface problems without any invocation of
an artificial periodicity.
In this method a finite voltage may occur between the two interstitial boundaries. 
It seems that this treatment of the potential step feature via
plane-waves produces thicker interstitial regions than necessary.

%
%

This work presents the general solution of the Poisson equation for
two dimensional periodic systems  based on the multipole compensation
method for a scheme of overlapping localized site densities. 
In the compact site density ansatz (see Refs. \cite{FPLO1,FPLO2}) the
total electronic charge density is written as a locally finite sum
without any interstitial density. The presented derivation differs from
what is given by Weinert \cite{Weinert} or Bertaut \cite{Bertaut} for 3D
periodic systems. The correct capacitor-like boundary condition is taken
explicitly into account, which results in a direct relation between the
finite voltage and the dipole of the slab as is expected from macroscopic
electrostatics of a film geometry.

In Appendix \ref{1dperiodicity} we shortly sketch how our ideas apply
to the 1D periodic problem, which differs in methodology from the
treatments\cite{Langridge,Delhalle,Porto} presented elsewhere, however
resulting in the same expressions.
%
%
\section{The multipole compensation in the localized site density scheme}
\begin{figure}[h!]
\begin{center}
\rotatebox{0}{\resizebox{75mm}{!}{\includegraphics{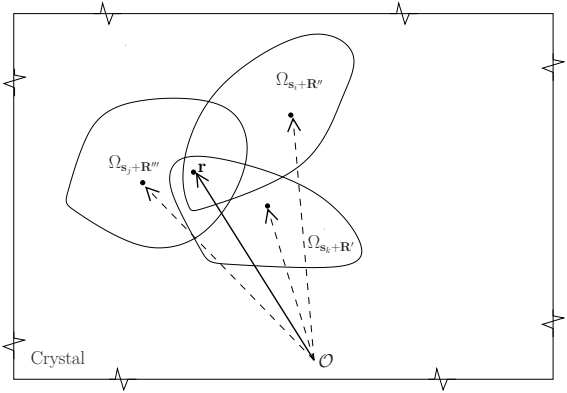}}}
\caption{\label{chargeoverlap}
Schematic site charge (compact support) distribution are overlapping at the
point $\mathbf{r}$.} 
\end{center}
\end{figure}

For crystalline structures in the thermodynamic limit there is an
infinite periodic adjustment of 
the finite  number of sites in the unit cell $\{\mathbf{s}_1,\ldots,\mathbf{s}_d\}$,
by the Bravais vectors $\mathbf{R}$. In the case of film geometry the
Bravais vectors $\mathbf{R}$ span one of the 2D Bravais
lattices. In other words, the slab geometry means
a three dimensional object with two dimensional periodicity. The symmetry
groups of this geometry are called layer groups
\cite{CrystallographyE}. According to the main assumption at any point
$\mathbf{r}$ in the extended solid only a finite number of site charges
with compact support $\Omega_{\mathbf{s}_j}\ j\in\{1,\ldots,d\}$ will
overlap, as it is shown in Fig. \ref{chargeoverlap}. Each site domain
may contain several neighboring unit cells.
\\
%
%
Using the local expansion ansatz 
\begin{equation}
|\mathbf{k}n\rangle=\sum_{\mathbf{R}}\sum_{j}^d|\mathbf{R}\mathbf{s}_j\mathcal{L}\rangle 
c_{\mathcal{L}\mathbf{s}_j}^{\mathbf{k}n}e^{i\mathbf{k}(\mathbf{R}+\mathbf{s}_j)}
\label{localansatz}
\end{equation}
for the Bloch state $|\mathbf{k}n\rangle$, where $\mathbf{k}$ is the
crystal momentum, $n$ is the band index and $\mathcal{L}$ is a composite
index for atomic orbitals containing the principal and the angular
quantum numbers, a local expression can be derived for the total density
$\rho$ 
\begin{eqnarray}
\rho(\mathbf{r})=\sum_{\mathbf{k},n}^{\text{occ}}\langle\mathbf{r}|\mathbf{k}n\rangle\langle\mathbf{k}n|
\mathbf{r}\rangle=
\sum_{\mathbf{R}}\sum_{j=1}^d \rho_j(\mathbf{r}-\mathbf{R}-\mathbf{s}_j),
\label{localdensity}
\end{eqnarray}
which defines the site-density $\rho_j$ and the corresponding domain
$\Omega_{\mathbf{s}_j+\mathbf{R}}$, see Koepernik and Eschrig \cite{FPLO1}.
%
%
The total charge density
\begin{multline}
\rho(\mathbf{r})=\sum_{\mathbf{R},j}^{\mathbf{r}\in\Omega_{\mathbf{R}+\mathbf{s}_j}}\rho_j(\mathbf{r}-\mathbf{R}-\mathbf{s}_j)=\\
\sum_{\mathbf{R},j}^{\mathbf{r}\in\Omega_{\mathbf{R}+\mathbf{s}_j}}\sum_{L}\rho_{jL}(|\mathbf{r}-\mathbf{R}-\mathbf{s}_j|)Y_{L}
(\mathbf{r}-\mathbf{R}-\mathbf{s}_j),
\end{multline}
is accordingly a finite sum of site contributions and is written as a
uniformly convergent sum of spherical harmonics 
$Y_{L},\ L=(\ell,m)$. The first line defines the total density
by the site densities. The complicated restriction in the summation
allows only such $\mathbf{R}$ and $j$ indices that the domain of
the corresponding site density contains the point $\mathbf{r}$. The
lattice periodicity of the total electronic density $\rho({\mathbf{r}})$
is not broken because of the $\mathbf{r}$ dependence of the summation.
%

To calculate the electrostatic potential $\phi(\mathbf{r})$ of the system one
needs to solve the Poisson equation
\begin{equation}
\vartriangle \phi(\mathbf{r})=\sum_{\mathbf{R},j}^{\mathbf{r}\in\Omega_{\mathbf{R}+\mathbf{s}_j}}\rho_j(\mathbf{r}-\mathbf{R}-\mathbf{s}_j)-
\sum_{\mathbf{R},j}Z_{j}\delta(\mathbf{r}-\mathbf{R}-\mathbf{s}_j)
\label{poissonequation}
\end{equation}
at any $\mathbf{r}\in\mathbbm{R}^3$, where no restriction is used in
the second Dirac delta summation. To get '{\it bulk}' quantities PBC is used in
the periodic directions (say $x$ and $y$) while in the third direction
the general finite voltage boundary condition is considered.
Trough the paper atomic units are used.
%
%

We now introduce the generalized Ewald densities expanded by spherical harmonics
\begin{eqnarray}\hspace{-0.3cm}&&
\rho^{\text{Ewald}}_{j}(\mathbf{r})=\sum_L\rho^{\text{Ewald}}_{jL}(r)Y_L(\mathbf{r})
\nonumber\\&&
\rho^{\text{Ewald}}_{jL}(r)=
A_{jL}\mathcal{N}_{\ell}r^{\ell}e^{-r^2p^2},\quad \mathcal{N}_{\ell}=
\frac{2p^{2\ell+3}}{\Gamma\left(\frac{2\ell+3}{2}\right)}\nonumber\\&&
\rho^{\text{Ewald}}_{j}(\mathbf{r})=\frac{2p^3}{\pi}\sum_L
\left((rp)^{\ell}e^{-r^2p^2}\right)\left(A_{jL}p^{\ell}\right)\frac{\pi Y_L(\mathbf{r})}{\Gamma(\ell+\frac{3}{2})}\nonumber\\,
\label{ewalddensity}
\end{eqnarray}
where $p$ is the usual Ewald parameter and the multipole freedom is settled in
$A_{jL}$. $\mathcal{N}_{\ell}$ is the normalization to have unity
multipole components for the Gaussian without $A_{jL}$. With that
normalization $A_{jL}$  are not dimensionless quantities, but their
dimensions are $[A_{jL}]=\text{meter}^{\ell}$. The last line of
Eq.(\ref{ewalddensity}) indicates the most reasonable coupling of terms
what is used in the calculations below. 
%
%
By the additional multipole compensators (Ewald charges) a modified
electrostatic problem is created with a new Poisson equation 
\begin{multline}
\vartriangle \tilde{\phi}(\mathbf{r})=\tilde{\rho}(\mathbf{r})\\
\tilde{\rho}(\mathbf{r})=
\sum_{\mathbf{R},j}^{\mathbf{r}\in\Omega_{\mathbf{R}+\mathbf{s}_j}}
\rho_j(\mathbf{r}-\mathbf{R}-\mathbf{s}_j)-
\sum_{\mathbf{R},j}Z_{j}\delta(\mathbf{r}-\mathbf{R}-\mathbf{s}_j)+\\
\sum_{\mathbf{R},j}\rho_{j}^{\text{Ewald}}(\mathbf{r}-\mathbf{R}-\mathbf{s}_j),
\label{modifiedpoissonequation}
\end{multline}
at any $\mathbf{r}\in\mathbbm{R}^3$. 
%
%
Taking the difference of the two Poisson
equations and using the linearity of the differential operator
$\vartriangle$, the Ewald problem 
\begin{multline}\hspace{-0.3cm}
\vartriangle\phi^{\text{Ewald}}(\mathbf{r})=\sum_{\mathbf{R},j}
\rho_{j}^{\text{Ewald}}(\mathbf{r}-\mathbf{R}-\mathbf{s}_j)=
\rho^{\text{Ewald}}(\mathbf{r}),
\label{ewaldequation}
\end{multline}
is derived. Thus, the complicated original Poisson problem is subdivided
into two simpler problems, 
$\phi(\mathbf{r})=\tilde{\phi}(\mathbf{r})-\phi^{\text{Ewald}}(\mathbf{r})$.
%
%
Before discussing the boundary conditions for the latter two
equations the solutions of the multipole compensations are needed because
they define the Ewald densities and thus the two Poisson equations
Eqs.(\ref{modifiedpoissonequation},\ref{ewaldequation}). Keeping the 2D periodicity in both subproblems is
compatible with the original PBC. 

A difficulty shows up in the solution of
Eq.(\ref{modifiedpoissonequation}) by the non-restricted sum of
overlapping non-compact Ewald densities. To get rid of this difficulty
one can assume that the Ewald densities are negligible outside the site domains.
This results the same finite number of terms in the last sum of
Eq.(\ref{modifiedpoissonequation}) as in the first sum. 
Using the approximate compactness of the Ewald densities the multipole
compensation equations are written as integrals over the whole space
\begin{multline}
0= \int_{\mathbbm{R}^3} d^3r\ Y_{\ell m}^{\ast}(\mathbf{r})r^{\ell}\Big(
\rho_j(\mathbf{r}-\mathbf{R}-\mathbf{s}_j)-Z_j\delta(\mathbf{r}-\mathbf{R}-\mathbf{s}_j)+\\
\rho^{\text{Ewald}}_{j}(\mathbf{r}-\mathbf{R}-\mathbf{s}_j)\Big).
\label{multipolecompensation}
\end{multline}
The introduced assumption simplifies the Poisson equation
Eq.(\ref{modifiedpoissonequation}) into the form 
\begin{eqnarray}&&
\vartriangle\tilde{\phi}(\mathbf{r})\approx\bar{\rho}(\mathbf{r}),
\label{newmodifiedpoissonequation}
\end{eqnarray}
where the r.h.s. is given as an $\mathbf{r}$-dependent {\it finite} sum of
site densities. This equation is referred below as the short-range
Poisson equation. The accuracy of the
approximation is measured by the localization of the applied Ewald
densities.
%

The linearity of $\vartriangle$ allows one to look for
the solution also in a finite sum
\begin{equation}
\tilde{\phi}(\mathbf{r})=\sum_{\mathbf{R}j}^{\mathbf{r}\in\Omega_{\mathbf{R}+\mathbf{s}_j}}
\tilde{\phi}_j(\mathbf{r}-\mathbf{R}-\mathbf{s}_j),
\label{tildephiansatz}
\end{equation}
at any point $\mathbf{r}$, where $\tilde{\phi}_j$ has the same compact
support as the site density $\rho_j$. Accordingly on each site  domain
$\Omega_{\mathbf{s}_j}$ a Dirichlet boundary problem is derived by the
multipole compensations. Extrapolating these boundary conditions to the
discussed equation Eq.(\ref{modifiedpoissonequation}) gives that $\tilde{\phi}$
has to vanish at the boundary of a composite compact region. Of course,
the in-plane periodicity is kept by means of the $\mathbf{r}$-dependent
summation. The solution of a site boundary problem with
vanishing potential at the boundaries is not discussed in the paper
because it corresponds with the 3D periodic case which is already published
in Refs. \cite{Weinert,FPLO1}.
%
%

Accordingly, the Ewald problem shows also the obvious infinite lattice
periodicity and carries the finite voltage boundary condition in the surface normal
direction. 
The infinite lattice periodicity allows one to use Fourier
expansion for the functions 
\begin{eqnarray}
\rho^{\text{Ewald}}(\mathbf{r})
=\sum_{\mathbf{K}_{||}}e^{i\mathbf{K}_{||}\mathbf{r}}\rho^{\text{Ewald}}(\mathbf{K}_{||},z) \nonumber \\
\phi^{\text{Ewald}}(\mathbf{r})=\sum_{\mathbf{K}_{||}}e^{i\mathbf{K}_{||}\mathbf{r}}\phi^{\text{Ewald}}(\mathbf{K}_{||},z) 
\label{fourierexpansion}
\end{eqnarray}
and the Ewald problem Eq.(\ref{ewaldequation}) is transformed into the reciprocal $\mathbf{K}_{||}$ space
\begin{multline}
\frac{d^2}{dz^2}\phi^{\text{Ewald}}(\mathbf{K}_{||},z) -K_{||}^2\phi^{\text{Ewald}}(\mathbf{K}_{||},z) 
=\rho^{\text{Ewald}}(\mathbf{K}_{||},z) \\
\rho^{\text{Ewald}}(\mathbf{K}_{||},z) =\frac{1}{\mathcal{U}}\sum_{j=1}^d\int_{\mathbbm{R}^2}d^2r \ 
\rho^{\text{Ewald}}_{j}(\mathbf{r}-\mathbf{s}_j)e^{-i\mathbf{K}_{||}\mathbf{r}},
\label{mainequation}
\end{multline}
where $\mathcal{U}$ means the volume of the 2D unit cell and
$d^2r$ denotes $dxdy$. The structure of the last equation
dictates different treatment for $K_{||}=0$ and $K_{||}\ne 0$. Since
the $K_{||}=0$ Fourier coefficient is a constant in-plane function at
any $z$ value the $K_{||}=0$ equation provides the required finite
voltage boundary condition whereas all the $K_{||}\ne 0$ solutions
vanish at $z=\pm\infty$. 
%
%
\section{Solution of the Ewald problem}
The Ewald boundary problem is defined in the previous section with two
subproblems. As mentioned previously the $K_{||}=0$ case differs notably
from the ordinary $K_{||}\ne 0$ case and thus needs a different
treatment. The $K_{||}\ne 0$ case contains a
non-singular differential operator on the l.h.s. which ensures
applicability of the Green function technique in the solution. On the
other hand the $K_{||}=0$ case can be solved by direct integration starting from
the charge neutrality condition and using the given boundary conditions.

\subsection{Solution of Eq.(\ref{mainequation}) for $\mathbf{K}_{||}\ne0$}
\addtocounter{section}{1}
This equation is a so called 1D modified Helmholtz equation and its
Green function reads
\begin{eqnarray}&&
G(z,\tilde{z})=-\frac{e^{-|z-\tilde{z}|K_{||}}}{2K_{||}},
\label{greenfunction}
\end{eqnarray}
including the vanishing boundary conditions at $\pm\infty$, which
results a complicated integral for the solution
\begin{multline}
\phi^{\text{Ewald}}(\mathbf{K}_{||},z)= \\
-\sum_j\frac{e^{-i\mathbf{K}_{||}\mathbf{s}_j}}{\mathcal{U}}
\int_{\mathbbm{R}^3}d^3\tilde{\mathbf{r}}\ 
\left(\frac{e^{-|z-s_{jz}-\tilde{z}|K_{||}}}{2K_{||}}\right)\times\\
\rho^{\text{Ewald}}_j(\tilde{\mathbf{r}})\ e^{-i\mathbf{K}_{||}\tilde{\mathbf{r}}}.
\label{fourierewaldpotential}
\end{multline}
The difficulty arises with the spherical harmonics expansion of
$\rho^{\text{Ewald}}_{j}$, which is the most reasonable choice to
have a straightforward solution of the short-range Poisson problem. Here
this advantage turns into a troublesome disadvantage because
the Green function and the exponential
$e^{-i\mathbf{K}_{||}\tilde{\mathbf{r}}}$ are preferably handled in
a Cartesian coordinate system.
%
%

With the help of the Fourier transform of the Green function
corresponding to the $z-s_{jz}$ variable ,
\begin{multline}
-\frac{1}{\sqrt{2\pi}}\int_{-\infty}^{+\infty}d\omega\ e^{i\omega(z-s_{jz})}F(\omega,\tilde{z})=-\frac{e^{-|z-s_{jz}-\tilde{z}|K_{||}}}{2K_{||}},
\\
F(\omega,\tilde{z})=
\frac{e^{-i\omega\tilde{z}}}{\sqrt{2\pi} (\omega^2+K_{||}^2)}
\end{multline}
one can introduce an $e^{-i\omega\tilde{z}}$ function which can be
coupled to the $e^{-i\mathbf{K}_{||}\tilde{\mathbf{r}}}$ function. The
introduction of the new three dimensional vector 
\begin{eqnarray}&&
\mathbf{K}(\omega)=\begin{pmatrix}\mathbf{K}_{||}\\ \omega \end{pmatrix}
\label{newKvector2D}
\end{eqnarray}
allows one to make use of the Rayleigh relation
\begin{eqnarray}
e^{-i\mathbf{K}(\omega)\tilde{\mathbf{r}}}=
4\pi\sum_{L'}(-i)^{\ell'}Y_{L'}(\mathbf{K}(\omega))j_{\ell'}(K(\omega)\tilde{r})Y^{\ast}_{L'}(\tilde{\mathbf{r}}),
\end{eqnarray}
where $j_{\ell}$ are the spherical Bessel functions. In this way the
problem is transformed into a form that is similar to the 3D periodic
case with the only difference that now the third reciprocal direction
$\omega$ is continuous. 

Interchanging the order of the Fourier $\int d\omega$ and the real space
integration $\int d^3\tilde{\mathbf{r}}$ and doing a  straightforward
but lengthy calculation the following equation is resulted 
\begin{multline}
\phi^{\text{Ewald}}(\mathbf{K}_{||},z) =
-\sum_{jL}\frac{e^{-i\mathbf{K}_{||}\mathbf{s}_j}}{\mathcal{U}}
\frac{(-i)^{\ell}}{\sqrt{\pi}}\left(e^{-\frac{K_{||}^2}{4p^2}}A_{jL}p^{\ell}\right)
\times\\
\int_{\mathbbm{R}}d\omega \ \frac{\pi Y_{L}(\begin{pmatrix}\mathbf{K}_{||}\\ \omega\end{pmatrix})}{\Gamma(\ell+\frac{3}{2})}\ 
\frac{(K_{||}^2+\omega^2)^{\frac{\ell-2}{2}}}{(2p)^{\ell}}e^{-\omega^2/(4p^2)}e^{i\omega(z-s_{jz})}.
\label{mainequation2}
\end{multline}
Here, the prefactor in the first brackets is dimensionless and the
$\omega$-integral shows the dimension of length. In the calculation of
the $\omega$-integral a representation of the spherical harmonics by
hypergeometric functions is used. An appropriate choice provides not
only a simplified writing of the solution but also a representation that
allows an accurate numerical implementation. 
It is worth to mention that taking any
representation of $Y_L$ in Eq.(\ref{mainequation2}) the $L=(0,0)$ case
will result the same integral, this case is come down from
the Fourier transform of the Green function.
According to the handy relations between the spherical angle-coordinates $(\vartheta,\varphi)=\hat{\mathbf{r}}$ and
the variables $\mathbf{K}_{||}$ and $\omega$ 
 \begin{eqnarray}&&
\sin{\vartheta}=\frac{K_{||}}{\sqrt{K_{||}^2+\omega^2}},\quad
\cot{\vartheta}=\frac{\omega}{K_{||}},\nonumber\\&&
e^{im\varphi}=\frac{(K_{||x}+iK_{||y})^m}{K_{||}^m}
\label{anglerelations}
\end{eqnarray}
the most reasonable choice is given by 
\begin{multline}
-\frac{2^{\ell+1}\sqrt{\pi}Y_L(\hat{\mathbf{r}})}{(2\ell+1)!!}=\frac{C(\ell,m)}{2^{\ell}}e^{im\varphi}(\sin{\vartheta})^{\ell}
\ \times\\ 
\left\{
\begin{array}{l}
_2F_1(-\frac{\ell+m}{2},-\frac{\ell-m}{2};-\frac{2\ell-1}{2};\frac{1}{\sin^2{\vartheta}})\\[1.5mm]
\text{\hspace{2cm}if\ } \ell+m \text{\ is even}\\\\
\cot{\vartheta}\ F(-\frac{\ell+m-1}{2},-\frac{\ell-m-1}{2};-\frac{2\ell-1}{2};\frac{1}{\sin^2{\vartheta}})\\[1.5mm]
\text{\hspace{2cm}if\ } \ell+m \text{\ is odd}\\
\end{array}\right.
\label{sphericalbyhypergeometric1}
\end{multline}
(see Varshalovich \cite{Varshalovich} , page 137, Eq.(30))
with the real valued quantity
\begin{eqnarray} &&\hspace{-5mm}
C(\ell,m)=
\frac{2^{\ell}(-1)^{\lfloor{\frac{\ell+m}{2}\rfloor+1}}}{\sqrt{(l+m)!(l-m)!(2\ell+1)}},
\end{eqnarray}
where $\lfloor\cdots\rfloor$ indicates the greatest integer function or floor.
$_2F_1$ is the hypergeometric function and $(\cdots)!!$ denotes the double factorial defined by
\begin{eqnarray}
n!!=\left\{\begin{array}{ccc}
n\cdot (n-2)\ldots 5\cdot 3\cdot 1 & n>0\ ,&\ \text{\ odd} \\
n\cdot (n-2)\ldots 6\cdot 4\cdot 2 & n<0\ ,&\ \text{\ even}\\
1&\multicolumn{2}{c}{n=-1,0}
\end{array} \right. .
\end{eqnarray}
In Eq.(\ref{mainequation2}) the $A_{jL}$ quantities are calculated by
means of  the multipole compensation. Experience shows that the quantity
$A_{jL}$ decays quickly with higher $\ell$ values. This fact enables one
to apply a cut-off in the summation which usually has a value of
$\ell_{\text{max}}=12$. 

For any finite $\ell$ the hypergeometric functions above terminate after
finite number of terms  
\begin{equation}
_2F_1^{\text{e/o}}=\sum_{k=0}^{k_{\text{max}}(\ell,m)}\alpha_k^{\text{e/o}}(\ell,m) K_{||}^{-2k} (K_{||}^2+\omega^2)^k,
\label{simplifiedF}
\end{equation}
because one of the first two arguments of $F$ is always a non-positive
integer, $k>0$. Here the subscript e/o distinguishes the even and odd values
of $(\ell+m)$. The expansion coefficients
$\alpha_k^{\text{e/o}}(\ell,m)$ are real with alternating
sign and $k_{\text{max}}(\ell,m)$ is the $\ell$ and $m$ dependent
termination  
value, see Table \ref{kmaxtable}.
\begin{table}[h!]
\begin{tabular}{|c|c|c|c|c|c|c|c|}
\hline
$(\ell,m)$ & $k_{\text{max}}$ & 
$(\ell,m)$ & $k_{\text{max}}$ & 
$(\ell,m)$ & $k_{\text{max}}$ &
$(\ell,m)$ & $k_{\text{max}}$ \\
\hline\hline
$(0,0) $ & 0 & $(3,-3)$ & 0 & $(4,-2)$ & 1 & $(5,-3)$ & 1\\
$(1,-1)$ & 0 & $(3,-2)$ & 0 & $(4,-1)$ & 1 & $(5,-2)$ & 1\\
$(1,0) $ & 0 & $(3,-1)$ & 1 & $(4,0) $ & 2 & $(5,-1)$ & 2\\
$(1,1) $ & 0 & $(3,0) $ & 1 & $(4,1) $ & 1 & $(5,0)$  & 2\\
$(2,-2)$ & 0 & $(3,1 )$ & 1 & $(4,2) $ & 1 & $(5,1)$  & 2\\
$(2,-1)$ & 0 & $(3,2) $ & 0 & $(4,3) $ & 0 & $(5,2)$  & 1\\
$(2,0) $ & 1 & $(3,3) $ & 0 & $(4,4) $ & 0 & $(5,3)$  & 1\\
$(2,1) $ & 0 & $(4,-4)$ & 0 & $(5,-5)$ & 0 & $(5,4)$  & 0\\
$(2,2) $ & 0 & $(4,-3)$ & 0 & $(5,-4)$ & 0 & $(5,5)$  & 0\\
\hline\hline
\end{tabular}
\caption{\label{kmaxtable}
The $k_{\text{max}}(\ell,m)$ values for $\ell \in \{1,\ldots,5\}$.}
\end{table}\\
Henceforth the even and odd contributions are treated separately. The
$\omega$-integrals from Eq.(\ref{mainequation2}) with the
third bracketed prefactor are written as
\begin{eqnarray}&&
\mathcal{I}^{\text{e}}=C(\ell,m)e^{im\varphi} P_{\ell m}K_{||}
\int_{\mathbbm{R}}d\omega\ \frac{_2F_1^{\text{e}}e^{-\omega^2/(4p^2)}}{(K_{||}^2+\omega^2)}e^{i\omega(z-s_{jz})},\nonumber\\&&
\mathcal{I}^{\text{o}}=C(\ell,m)e^{im\varphi}
P_{\ell m}\ 
\int_{\mathbbm{R}}d\omega\
\omega\frac{_2F_1^{\text{o}}e^{-\omega^2/(4p^2)}}{(K_{||}^2+\omega^2)}e^{i\omega(z-s_{jz})},\nonumber\\&&
P_{\ell m}=\frac{e^{-K_{||}^2/(4p^2)}}{2p}\left(\frac{K_{||}}{2p}\right)^{\ell-1}(A_{jL}p^{\ell}) .
\label{evenoddintegrals}
\end{eqnarray}
Inserting Eq.(\ref{simplifiedF}) into the above integrals results the
finite series
\begin{multline}
\frac{_2F_1^{\text{e/o}}e^{-\omega^2/(4p^2)}}{(K_{||}^2+\omega^2)}=
\frac{e^{-\omega^2/(4p^2)}}{(K_{||}^2+\omega^2)}\ +\\
\sum_{n=0}^{k_{\text{max}}(\ell,m)-1}
K_{||}^{-2-2n}
\omega^{2n}e^{-\omega^2/(4p^2)}
\mathcal{G}^{\text{e/o}}(n,\ell,m)
\label{omegaintegrand}
\end{multline}
in the integrands, where the $n$-summation comes from the binomial
expansion of $(K_{||}^2+\omega^2)^k$ with $n\ge 0$ and
\begin{eqnarray}&&
\mathcal{G}^{\text{e/o}}(n,\ell,m)=\sum_{k=n}^{k_{\text{max}}(\ell,m)-1}
\alpha_{k+1}^{\text{e/o}}(\ell,m)\binom{k}{n}.
\label{gevenodd}
\end{eqnarray}
To derive Eq.(\ref{omegaintegrand}) the order of the finite $k$ and $n$ summations
is interchanged. After substituting the last two equations into
Eq.(\ref{evenoddintegrals}) a convenient handling,
\begin{eqnarray}&&
\mathcal{I}^{\text{e}}=C(\ell,m)e^{im\varphi}P_{\ell m}\ \times\nonumber\\&& \left(
\mathcal{I}_3(z,j,K_{||})\ +
\sum_{n=0}^{k_{\text{max}}(\ell,m)-1}\tilde{\mathcal{I}}_{2n}\ 
\ \mathcal{G}^{\text{e}}(n,L)
 \right)\nonumber\\&&
\mathcal{I}^{\text{o}}=C(\ell,m)e^{im\varphi}P_{\ell m}\ \times\nonumber\\&& \left(
\mathcal{I}_4(z,j,K_{||})\ +
\sum_{n=0}^{k_{\text{max}}(\ell,m)-1}\tilde{\mathcal{I}}_{2n+1}\ 
\ \mathcal{G}^{\text{o}}(n,L).\right)
\label{Ioddintegral}
\end{eqnarray}
is achieved for the lengthy expressions by introducing three integrals
$\mathcal{I}_3, \mathcal{I}_4$ and $\tilde{\mathcal{I}}_{\gamma}$, where
$\gamma$ can be even or odd. Taking the values of $k_{\text{max}}$ from
Table \ref{kmaxtable}  it turns out that in the cases of
$k_{\text{max}}(\ell,m)=0$ only $\mathcal{I}_3$ and $\mathcal{I}_4$
provide contributions. These two integrals are defined as
\begin{eqnarray}&&
\mathcal{I}_3(z,j,K_{||})=
\int_{\mathbbm{R}} dx\ 
\frac{e^{-K_{||}^2x^2/(4p^2)}}{1+x^2}
e^{iK_{||}(z-s_{jz})x} \nonumber\\&&
\mathcal{I}_4(z,j,K_{||})=
\int_{\mathbbm{R}} dx\ 
\frac{xe^{-K_{||}^2x^2/(4p^2)}}{1+x^2}
e^{iK_{||}(z-s_{jz})x},
\nonumber\\
\label{fourintegrals2}
\end{eqnarray}
where $\omega=K_{||}x$ is used. They look like a Fourier
transformation which suggests the use of the convolution theorem of Fourier
transformations. The final results of the integrals are presented in
Appendix \ref{appendixintegral1}. The $\tilde{\mathcal{I}}_{2n}$ and $\tilde{\mathcal{I}}_{2n+1}$ integrals are
treated together by introducing the dimensionless new variable $y$
\begin{eqnarray}&&
\tilde{\mathcal{I}}_{\gamma}(z,j,K_{||})=\nonumber\\&&
2^{\gamma+1}
\int dy\ y^{\gamma}e^{-(K_{||}^2y^2/p^2)+2(iK_{||}(z-s_{jz}))y}=\nonumber\\&&
i^{\gamma}
\left(\frac{2p}{K_{||}}\right)^{\gamma +1}\ 
\sum_{t=0}^{\lfloor{\frac{\gamma}{2}\rfloor}}
\left(\binom{\gamma}{2t}\Gamma (t+\frac{1}{2})(-1)^t\right)
\ \times\nonumber\\&&
e^{-p^2(z-s_{jz})^2}(p(z-s_{jz}))^{\gamma -2t},
\label{fourintegrals1}
\end{eqnarray}
where $\omega=2K_{||}y$ and $\ell-2\ge \gamma$. The factor in the
first bracket describes the $K_{||}$ dependence and the exponential from
$P_{\ell m}$ ensures its decaying behavior for large $K_{||}$. It is also worth
to mention that the parity of $\gamma$ completely separates the
results. For even $\gamma$ the integral has a real value while for the odd
cases it is imaginary.

%
%

From the numerical numerical point of view in the above sums the 
the quickly vanishing $A_{jL}$s determine the order of the contributions
and  ensure the inclusion of all relevant contributions.

A partial simplification can be achieved in the $\mathcal{I}^{\text{e/o}}$ calculation
if the order of the $n$- and $t$-summation is interchanged.
Hereby the $n$-summation with the $(z-s_{jz})$ functions goes ahead and
the same upper limits can be used in the summations,
\begin{widetext}
\begin{eqnarray}
&\mathcal{I}^{\text{e}}=&
C(\ell,m)e^{im\varphi}P_{\ell m}\mathcal{I}_3+\nonumber\\&&
C(\ell,m)e^{im\varphi}P_{\ell m}
\sum_{n=0}^{k_{\text{max}}-1}
(-1)^ne^{-p^2(z-s_{jz})^2}(p(z-s_{jz}))^{2n}
\sum_{t=n}^{k_{\text{max}}-1}
\left(\frac{2p}{K_{||}}\right)^{2t+1}
\left(\binom{2t}{2n}\Gamma (t-n+\frac{1}{2})\right)
\mathcal{G}^{\text{e}}(t,L);
\nonumber\\
&\mathcal{I}^{\text{o}}=&
C(\ell,m)e^{im\varphi}P_{\ell m}\mathcal{I}_4+\nonumber\\&&
iC(\ell,m)e^{im\varphi}P_{\ell m}
\sum_{n=0}^{k_{\text{max}}-1}
(-1)^ne^{-p^2(z-s_{jz})^2}(p(z-s_{jz}))^{2n+1}
\sum_{t=n}^{k_{\text{max}}-1}
\left(\frac{2p}{K_{||}}\right)^{2t+2} 
\left(\binom{2t+1}{2n+1}\Gamma (t-n+\frac{1}{2})\right)
\mathcal{G}^{\text{o}}(t,L).\nonumber\\
\end{eqnarray}
\end{widetext}

The final real valued $K_{||}\ne 0$ Ewald potential can be easily
derived from the result
\begin{eqnarray}&&
\phi^{\text{Ewald}}_{\ne
  0}(\mathbf{r})=\sum_j{\sum_{{K}_{||}}}'\sum_L
\left(e^{i\mathbf{K}_{\parallel}(\mathbf{r}-\mathbf{s}_j)}e^{-i\ell\pi/2}e^{i
  m\varphi}\right)\times\nonumber\\&&
\frac{C(\ell,m)}{\mathcal{U}\sqrt{\pi}}P_{\ell m}\left\{
\begin{array}{ll}
\begin{pmatrix}\mathcal{I}_3+\sum_n\tilde{\mathcal{I}}_{2n}\mathcal{G}^{\text{e}}\\0\end{pmatrix}
& \ell+m\text{\ even}\\ \\
\begin{pmatrix}0\\
-i\mathcal{I}_4-i\sum_n\tilde{\mathcal{I}}_{2n+1}\mathcal{G}^{\text{o}}\end{pmatrix}
&  \ell+m\text{\ odd}
\end{array}\right.\nonumber\\&&
\label{realpartKne0}
\end{eqnarray}
where the two dimensional column vector denotes the real and imaginary
components of an arbitrary complex number. 

Simpler expressions can be found for the special cases if
$(z-s_{jz})=0$. It is easily established by symmetry reasons
that only for even $\gamma$ gives the integral
$\tilde{\mathcal{I}}_{\gamma}$ (see
Eq.(\ref{fourintegrals1})) a non-zero result
\begin{equation}
\tilde{\mathcal{I}}_{\gamma}(z=s_{jz})=
\left(\frac{K}{2p}\right)^{-\gamma-1}
\frac{\Gamma(\frac{\gamma}{2}+\frac{1}{2})}{4}.
\end{equation}
For the other two integrals ($\mathcal{I}_3,\mathcal{I}_4$) the results
can be calculated by using the general expressions given in the
Appendix \ref{appendixintegral1} (see also Gradshteyn\cite{Gradshteyn}, 3.466/1.) 
\begin{eqnarray}&&
\mathcal{I}_3(z=s_{jz})=\pi 
e^{K_{||}^2/(4p^2)}
\text{Erfc}\left(\frac{K_{||}}{2p}\right),\nonumber\\&&
\mathcal{I}_4(z=s_{jz})=0,
\label{integral2}
\end{eqnarray}
which tells that only the even $(\ell+m)$ terms contribute.

Easy to check that all the integral results vanish at the boundaries $z=\pm\infty$
which ensures that the final formula fulfills the required boundary
conditions. 
\subsection{Solution of Eq.(\ref{mainequation}) for $\mathbf{K}_{||}=0$}
In this section the solution of the boundary problem
\begin{eqnarray}&&
\frac{d^2}{dz^2}\phi^{\text{Ewald}}(0,z) =\rho^{\text{Ewald}}(0,z),\nonumber\\&&
\phi^{\text{Ewald}}(0,+\infty)-\phi^{\text{Ewald}}(0,-\infty)=C_2
\label{ewaldKnull} 
\end{eqnarray}
is looked for, where the Ewald densities are defined through the multipole
compensation and $C_2$ refers to the already introduced finite voltage. Taking
only the monopole equation 
\begin{equation}
\int _{\mathbbm{R}^3} d^3\mathbf{r} \ 
\Big(\rho_j(\mathbf{r}-{\mathbf{s}_j}-\mathbf{R})+
\rho^{\text{Ewald}}_{j}(\mathbf{r}-\mathbf{s}_j-\mathbf{R})\Big)-Z_j=0
\end{equation}
together with the charge neutrality equation
\begin{equation}
\sum_{\mathbf{R},j}\int_{-\infty}^{+\infty}dz\int _{\mathcal{U}} dxdy \
\rho_j(\mathbf{r}-\mathbf{s}_j-\mathbf{R})-\sum_jZ_j=0
\label{chargeneutrality}
\end{equation}
the well-known fact of the neutrality of the Ewald density lattice can be derived
\begin{equation}
\sum_j^d\int _{\mathbbm{R}^3} d^3\mathbf{r} \
\rho^{\text{Ewald}}_{j}(\mathbf{r}-\mathbf{s}_j)=
\sqrt{\frac{1}{4\pi}}\sum_j^dA_{j(0,0)}=0.
\label{ewaldneutrality}
\end{equation}
For a 2D periodic system Eq.(\ref{ewaldneutrality}) has the
consequence 
\begin{equation}
\int_{-\infty}^{+\infty}dz\ \alpha\rho^{\text{Ewald}}(K_{||}=0,z)=0,
\quad \alpha\in\mathbbm{R}\setminus\{0\}
\label{Knullrequirement}
\end{equation}
which can accurately be approximated by a bounded integration range
$[D_{\text{low}},D_{\text{up}}]$ for well localized Ewald charges. 
In words, one uses the bounded region where the Ewald charge
distribution is non-zero.
Integrating Eq.(\ref{ewaldKnull}) and applying the previous equation
results
\begin{eqnarray}&&
\int_{D_{\text{low}}}^{D_{\text{up}}}dz\ \alpha\left(\frac{d^2}{dz^2}\ \phi^{\text{Ewald}}(0,z)\right)\approx 0.
\end{eqnarray}
which yields the boundary behavior of the negative electric field component $E(z)=d\phi^{\text{Ewald}}(0,z)/dz$
\begin{equation}
\int_{D_{\text{low}}}^{D_{\text{up}}}dz\ \alpha\frac{dE(z)}{dz}=\alpha\left(E(D_{\text{up}})-E(D_{\text{low}})\right)\approx 0.
\end{equation}
Applying the Gauss law outside of this bounded region results the boundary problem
for $E$
\begin{equation}
\frac{dE(z)}{dz}=\rho^{\text{Ewald}}(0,z),\quad 
\begin{array}{l}
E(z\le D_{\text{low}})=\text{const.}\\
E(z\ge D_{\text{up}})=\text{const.},
\end{array}
\label{boundaryproblem_q}
\end{equation}
which defines another first order boundary problem for the $K_{||}=0$ Ewald 
potential 
\begin{equation}
\frac{d\phi^{\text{Ewald}}(0,z)}{dz}=E(z),\ 
\begin{array}{l}
\phi^{\text{Ewald}}(0,z\le D_{\text{low}})=0\\
\phi^{\text{Ewald}}(0,z\ge D_{\text{up}})=C_2
\end{array}\, ,
\label{boundaryproblem_phi}
\end{equation}
The finite values at the potential boundaries dictate the const.=0
conditions for $E$. It is worth to mention that applying
non-zero constant boundary conditions for $E$ contradicts with the
constant behavior of the potential outside of the bounded range
$[D_{\text{low}},D_{\text{up}}]$. 

Both problems can be solved by simple integration which results the
final expression 
\begin{eqnarray}&&
\phi^{\text{Ewald}}(0,z)= \nonumber\\&&\frac{1}{2}\left(
\int_{D_{\text{low}}}^zdz'\ E(z')-\int_z^{D_{\text{up}}}dz'\ E(z')\right)+
\frac{C_2}{2}\nonumber\\&&
\hspace{2cm} D_{\text{low}}\le z \le D_{\text{up}}
\label{integrationKnenull}
\end{eqnarray}
with explicit inclusion of the potential difference $C_2$. 
%
\subsection{Evaluation of Eq.(\ref{integrationKnenull})}

Integrating the boundary problem in Eq.(\ref{boundaryproblem_q}) results
\begin{eqnarray}
E(z')=&&\frac{1}{\mathcal{U}}\sum_{jL}
\frac{1}{2}\left(\int_{-D}^{z'-s_{jz}}dz''-\int^{+D}_{z'-s_{jz}}dz''\right)\times\nonumber\\&&
\int_{\mathbbm{R}^2}dx''dy''
\rho_{jL}^{\text{Ewald}}(r'')Y_{L}(\mathbf{r}'')
\end{eqnarray}
with a common $D$, for which $D\to\infty$, and $E$ will satisfy the requirement given by
Eq.(\ref{Knullrequirement}). Unlike in the $K_{||}\ne 0$ case the above
integral allows one to use cylindrical coordinates $(\varrho,\varphi,z'')$.
Using the representation of the spherical harmonics (see Varshalovich
\cite{Varshalovich}, page 137, Eq.(33)),
\begin{eqnarray}
Y_L(\hat{\mathbf{r}})=\zeta_{m0}e^{im\varphi}
\sqrt{\frac{2\ell+1}{4\pi}\frac{(\ell+|m|)!}{(\ell-|m|)!}}(\cos{\vartheta})^{\ell}
\frac{(\tan{\vartheta})^{|m|}}{|m|!2^{|m|}}\nonumber\\
\times F(-\frac{\ell-|m|}{2},-\frac{\ell-|m|-1}{2};|m|+1;-\tan^2{\vartheta})\nonumber\\
\zeta_{m0}=\left\{ \begin{array}{cl}
(-1)^m& m>0\\1&m\le 0
\end{array}\right.\nonumber\\
\label{sphericalbyhypergeometric2}
\end{eqnarray}
immediately results $2\pi\delta_{m,0}$ for the $\varphi$-integral and
simplifies the expression for $E$ to
\begin{eqnarray}
E(z')=\frac{2\pi}{\mathcal{U}}\sum_{j\ell}\mathcal{N}_{\ell}A_{j(\ell,0)}
\sqrt{\frac{2\ell+1}{4\pi}}
\sum_{k=0}^{\lfloor\frac{\ell}{2}\rfloor}\tilde{\alpha}_k(\ell,0)(-1)^{k}\times\nonumber\\
\frac{1}{2}\left(\int_{-D}^{z'-s_{jz}}dz''-\int^{+D}_{z'-s_{jz}}dz''\right)
\ z''^{(\ell-2k)}e^{-z''^2p^2}\times\nonumber\\
\int_0^{+\infty}d\varrho\  
\varrho^{2k+1}e^{-\varrho^2p^2}.\nonumber\\
\end{eqnarray}
Here the hypergeometric function is written again by a finite sum with
coefficients $\tilde{\alpha}_k(\ell)$. After some algebraic manipulations
with finite $\ell$ summation $E$ is expressed as
\begin{eqnarray}&&
E(z')=
%
%
\sum_{j=1}^d\sum_{\lambda=0}^{\lfloor\frac{\ell_{\text{max}}}{2}\rfloor}q_{2\lambda,j}
\sum_{k=0}^{\lfloor\frac{\ell_{\text{max}}-2\lambda}{2}\rfloor}C(j,2\lambda+2k,k)+\nonumber\\&&
\sum_{j=1}^d\sum_{\lambda=0}^{\lfloor\frac{\ell_{\text{max}}-1}{2}\rfloor}q_{2\lambda+1,j}
\sum_{k=0}^{\lfloor\frac{\ell_{\text{max}}-(2\lambda+1)}{2}\rfloor}C(j,2\lambda+1+2k,k)\nonumber\\&&
q_{t,j}=\frac{1}{2}\left(\int_{-D}^{z'-s_{jz}}dz-\int_{z'-s_{jz}}^{+D}dz\right)p^3(zp)^te^{-z^2p^2},\nonumber\\&&
C(j,\ell,k)=
%
%
\frac{2A_{j,(\ell,0)}p^{\ell}}{p^2\mathcal{U}}\frac{(-1)^k}{\sqrt{2\ell+1}}\binom{\ell}{2k}
\frac{\Gamma\left(k+\frac{1}{2}\right)}{\Gamma\left(\ell+\frac{1}{2}\right)}.
\label{approximatedq}
\end{eqnarray}
Here, the explicit form of the hypergeometric coefficients
$\tilde{\alpha}$ is applied and $C(j,\ell,k)$ is defined as a
dimensionless quantity. The one dimensional integral has the dimension
of a surface charge $[q_{t,j}]=1/{\text{m}^2}$. In the expression above the $\lfloor\cdots\rfloor$
bracketing is used again to denote the floor function. 

If $t$ is odd ($t=2\lambda+1$) then
\begin{equation}
q_{2\lambda+1,j}=
\frac{-e^{-(z'-s_{jz})^2p^2}p^2}{2}\sum_{n=0}^{\lambda}
\frac{\lambda!((z'-s_{jz})p)^{2n}}{n!}
\label{q2lambdaplus1}
\end{equation}
(see Gradshteyn\cite{Gradshteyn}, 3.351/2.). For $z'=s_{jz}$ only
the $n=0$ term is non-zero. For even $t$ $(t=2\lambda)$ the result
\begin{equation}
q_{2\lambda,j}=
\left\{
\begin{array}{ll}
\lambda=0&q_{0,j}=\frac{\sqrt{\pi}p^2}{2}\text{Erf}((z'-s_{jz})p)\\[4mm]
\lambda>0&
q_{2\lambda,j}=f(\lambda)
q_{0,j}-\frac{p^2}{2}e^{-(z'-s_{jz})^2p^2}\ \times\\[1.5mm]
&\sum_{n=1}^{\lambda}\Big(f(\lambda)/f(n)\Big)\ 
((z'-s_{jz})p)^{2n-1}
\end{array}\right.
\label{q2lambda}
\end{equation}
is proven in Appendix \ref{evenqintegral} and expressed with the
help of the error function $\text{Erf}(\cdots)$ and a gamma function
\begin{equation}
f(\lambda)=
\frac{1}{\sqrt{\pi}}\Gamma\left(\lambda+\frac{1}{2}\right)=
\left\{
\begin{array}{lr}
1&\lambda=0\\
(2\lambda-1)!!/2^{\lambda}&\lambda>0
\end{array}\right.,
\end{equation}
which shows also the identity
\begin{equation}
\sum_{n=0}^{\lambda}f(n)\frac{\lambda!}{n!}=
2f(\lambda+1)
\label{fidentity}
\end{equation}
proven by induction to $\lambda$. This identity is used below in the
calculations of the boundary properties of E.
Eq.(\ref{q2lambda}) is also valid in the case of $z'=s_{jz}$. Further
algebraic manipulations on $E$ leads to the expression
\begin{widetext}
\begin{eqnarray}&&
E(z')=
\sum_{j=1}^{d}
\frac{\sqrt{\pi}p^2}{2}\text{Erf}((z'-s_{jz})p)P_{\text{o}}(j,p,0)
%
%
-\sum_{j=1}^{d}\frac{e^{-(z'-s_{jz})^2p^2}p^2}{2}
\sum_{n=0}^{\lfloor\frac{\ell_{\text{max}}-1}{2}\rfloor}P_{\text{e}}(j,p,n)((z'-s_{jz})p)^{2n}\nonumber\\&&
%
-\sum_{j=1}^{d}\frac{e^{-(z'-s_{jz})^2p^2}p^2}{2}
\sum_{n=1}^{\lfloor\frac{\ell_{\text{max}}}{2}\rfloor}P_{\text{o}}(j,p,n)((z'-s_{jz})p)^{2n-1},
%
\end{eqnarray}
where
\begin{eqnarray}&&
P_{\text{e}}(j,p,n)=
\frac{2}{n!}
\sum_{k=n}^{\lfloor\frac{\ell_{\text{max}}-1}{2}\rfloor}
\frac{2A_{j,(2k+1,0)}(p/2)^{2k+1}}{p^2\mathcal{U}}\frac{\sqrt{\pi}}{\sqrt{4k+3}}
\frac{\Gamma\left(2k+2\right)}{\Gamma\left(2k+\frac{3}{2}\right)}
\left(\sum_{\lambda=n}^k\frac{(-1)^{k-\lambda}}{k!}\frac{(2\lambda)!!}{(2\lambda+1)!!}\binom{k}{\lambda}\right);
\nonumber\\&&
P_{\text{o}}(j,p,n)=
\frac{1}{f(n)}
\sum_{k=n}^{\lfloor\frac{\ell_{\text{max}}}{2}\rfloor}
\frac{2A_{j,(2k,0)}(p/2)^{2k}}{p^2\mathcal{U}}\frac{\sqrt{\pi}}{\sqrt{4k+1}}
\frac{\Gamma\left(2k+1\right)}{\Gamma\left(2k+\frac{1}{2}\right)}
\left(\sum_{\lambda=n}^k\frac{(-1)^{k-\lambda}}{k!}\binom{k}{\lambda}\right);\nonumber\\&&
P_{\text{o}}(j,p,0)=C(j,0,0)=\frac{2A_{j(0,0)}}{p^2\mathcal{U}}.
\label{PeandPo}
\end{eqnarray}

The $k$ sums above are numerically feasible representations. Summing up
$P_{\text{o}}(j,p,0)$ from Eq.(\ref{PeandPo}) to all sites $j$ results
zero by the Ewald neutrality Eq.(\ref{ewaldneutrality}). This identity 
is used to check the boundary properties of $E$. A detailed
justification of the vanishing site sum of $P_{\text{o}}(j,p,0)$ starting from
Eq.(\ref{approximatedq}) can be found in Appendix
\ref{appendixPojp0}. The whole space integral of $E$ connects 
the boundary values of $\phi^{\text{Ewald}}(0,z)$,
\begin{eqnarray}&&
C_2=\int_{-\infty}^{+\infty}dz'\ E(z')=
\lim_{D\to +\infty}\sum_{j=1}^d\frac{\sqrt{\pi}p^2}{2}P_{\text{o}}(j,p,0)\int_{-D}^{+D}\text{Erf}((z'-s_{jz})p)dz'-\nonumber\\ &&
\frac{p\sqrt{\pi}}{2}\sum_{j=1}^{d}\sum_{\lambda=0}^{\lfloor\frac{\ell_{\text{max}}-1}{2}\rfloor}
\left(\sum_{n=0}^{\lambda}\frac{\lambda!}{n!}f(n)\right)\sum_{t=\lambda}^{\lfloor\frac{\ell_{\text{max}}-1}{2}\rfloor}
C(j,2t+1,t-\lambda)
=-\sqrt{4\pi}\sum_{j=1}^ds_{jz}A_{j(0,0)}
-\frac{p\sqrt{\pi}}{2}\sum_{j=1}^{d}C(j,1,0)=\nonumber\\&&
-\frac{\sqrt{4\pi}}{\mathcal{U}}\sum_{j=1}^ds_{jz}A_{j(0,0)}
-\frac{1}{\mathcal{U}}\sqrt{\frac{4\pi}{3}}\sum_{j=1}^dA_{j(1,0)}.
\label{asymptotics1}
\end{eqnarray}
\end{widetext}
and results the value of $C_2$.
For the third sum in the second line one can apply the identity
from Eq.(\ref{fidentity}) and the equation Eq.(\ref{asymptotics2}) from Appendix
\ref{appendixPojp0}.

The equation
\begin{equation}
\sum_{j=1}^d\int_{\mathbbm{R}^3}d^3r\ z \rho_j(\mathbf{r}-\mathbf{s}_j)=
\int_{\mathbbm{R}}dz\int_{\mathcal{U}}dxdy\ z\rho(\mathbf{r})
\end{equation}
derived by the fact that the Bravais vectors have zero $z$ components
can be used in the $z$ component dipole compensation. The resulted relation
\begin{eqnarray}&&
\int_{\mathbbm{R}}dz\int_{\mathcal{U}}dxdy\ z\rho(\mathbf{r})=
-\sqrt{\frac{4\pi}{3}}\sum_{j=1}^dA_{j(1,0)}
+\sum_{j=1}^ds_{jz}Z_j\nonumber\\&&-\sqrt{4\pi}\sum_{j=1}^ds_{jz}A_{j(0,0)}=
C_2\mathcal{U}+\sum_{j=1}^ds_{jz}Z_j
\label{dipolecompensation}
\end{eqnarray}
expresses the finite voltage $C_2$ as the $z$ component slab dipole
(surface, area) density and gives the expected macroscopic
electrostatics of the system. 

\begin{widetext}
Turning to the calculation of the potential by
Eq.(\ref{integrationKnenull}) one arrives to the final expression
\begin{eqnarray}&&
\phi^{\text{Ewald}}(0,z)=
\sum_{j=1}^{d}\frac{p}{2}P_{\text{o}}(j,p,0)\sqrt{\pi}((z-s_{jz})p)\text{Erf}((z-s_{jz})p)
+\sum_{j=1}^{d}\frac{p}{2}P_{\text{o}}(j,p,0)e^{-(z-s_{jz})^2p^2}\nonumber\\&&
-\sum_{j=1}^{d}\sum_{n=0}^{\lfloor\frac{\ell_{\text{max}}-1}{2}\rfloor}\Big(f(n)P_{\text{e}}(j,p,n)\Big)
\frac{p}{4}\sqrt{\pi}\text{Erf}((z-s_{jz})p)\nonumber\\&&
+\sum_{j=1}^{d}\sum_{n=1}^{\lfloor\frac{\ell_{\text{max}}-1}{2}\rfloor}\Big(f(n)P_{\text{e}}(j,p,n)\Big)
\frac{p}{4}e^{-(z-s_{jz})^2p^2}
\sum_{t=1}^n\frac{((z-s_{jz})p)^{2t-1}}{f(t)}\nonumber\\&&
+\sum_{j=1}^{d}\sum_{n=1}^{\lfloor\frac{\ell_{\text{max}}}{2}\rfloor}\Big((n-1)!P_{\text{o}}(j,p,n)^{}\Big)
\frac{p}{4}e^{-(z-s_{jz})^2p^2}
\sum_{t=0}^{n-1}\frac{((z-s_{jz})p)^{2t}}{t!}+\frac{C_2}{2}
\end{eqnarray}
\end{widetext}
A straightforward calculation leads to the observation that only the
first term and the third one in the second line have contribution at the
boundaries. The calculation of the boundary values yields the same
results what is already calculated 
in Eq.(\ref{asymptotics1}).
\section{summary}
%
%
The paper proposes a general approach to solve the Poisson equation for
two dimensional periodic systems. The method is based on the
pseudo-charge method \cite{Weinert} and uses a localized site density scheme\cite{FPLO1,FPLO2},
where the total charge density is written as a locally finite sum.
The bounded domains $\Omega_{\mathbf{s}_j}$ coincide
with the compact support of the local electronic site charge densities
$\rho_j(\mathbf{r}-\mathbf{s}_j)$. 
Going beyond the Ewald
technique, the approach not only compensates the site monopoles
but also all higher multipoles by the additional generalized Ewald site
densities. 
Solutions of the multipole compensation equations yield the multipole
freedoms and unambiguously define the Ewald densities of every site.

The proper boundary condition for the slab geometry, i.e. the finite voltage
at the infinite boundaries in the $z$-direction, results in the correct
macroscopic electrostatics.
The 
finite potential difference at the boundaries is expressed
by the 
 total dipole (surface, area) density. 
To our knowledge, the explicit treatment of the general slab boundary conditions presents
the most simple and direct way for self-consistent calculations of the
finite voltage. 
The expressions presented here, may easily be applied 
to the electrostatics of lattices with
point-like multipoles for molecular dynamical simulations.

In Appendix \ref{1dperiodicity} we shortly sketch how our ideas apply
to the 1D periodic problem, which differs in methodology from the
treatments\cite{Langridge,Delhalle,Porto} presented elsewhere, however
resulting in the same expressions.
%
%
%
 
\appendix
%
\section{Calculation of the $\mathcal{I}_3,\ \mathcal{I}_4$ integrals}
\label{appendixintegral1}
By the Fourier transformations 
\begin{equation}
\frac{1}{\sqrt{2\pi}}\int_{\mathbbm {R}}dx \  \frac{1}{1+x^2}e^{i\tau x}=
\sqrt{\frac{\pi}{2}}e^{-|\tau|}
\nonumber
\end{equation}
\begin{equation}
\frac{1}{\sqrt{2\pi}}\int_{\mathbbm {R}}dx  \ e^{-\left(\frac{K_{||}}{2p}\right)^2x^2}e^{i(\delta-\tau)x}=
\frac{p\sqrt{2}}{K_{||}}e^{-\frac{(\delta-\tau)^2p^2}{K_{||}^2}}
\nonumber
\end{equation}
\begin{equation}
\frac{1}{\sqrt{2\pi}}\int_{\mathbbm {R}}dx  \ xe^{-\left(\frac{K_{||}}{2p}\right)^2x^2}e^{i(\delta-\tau)x}=
\frac{i4p^3(\delta-\tau)}{K_{||}^3\sqrt{2}}e^{-\frac{(\delta-\tau)^2p^2}{K_{||}^2}}
\end{equation}
(see Gradshteyn \cite{Gradshteyn} 3.354/5., 3.323/2., 3.462/2.)
,where $\delta_j=K_{||}(z-s_{jz})$ and by the convolution theorem one can calculate the required integrals
\begin{eqnarray}
\mathcal{I}_3=&&
\frac{1}{\sqrt{2\pi}}\int_{\mathbbm{R}}d\tau \ \mathcal{F}[f](\tau)\mathcal{F}[g](\delta_j-\tau)=\mathcal{F}[fg](\delta_j)=\nonumber\\&&
\frac{\pi}{2}e^{K_{||}^2/(4p^2)}
\left(\mathcal{D}(\delta_j)+\mathcal{D}(-\delta_j)\right)\nonumber\\
\mathcal{I}_4=&&
i\frac{\pi}{2}e^{K_{||}^2/(4p^2)}\left(\mathcal{D}(-\delta_j)-\mathcal{D}(\delta_j)\right)
\end{eqnarray}
where
\begin{equation}
\mathcal{D}(\delta)=e^{\delta}\text{Erfc}\left(\frac{K_{||}}{2p}+\frac{p\delta}{K_{||}}\right).
\end{equation}
%
%
\section{Calculation of the $q_{2\lambda}$ integral}
\label{evenqintegral}
Easy to find out the recursive relation
\begin{equation}
q_{2\lambda,j}=-\frac{(z'-s_{jz})^{2\lambda-1}}{2p^2}e^{-(z'-s_{jz})^2p^2}+
\frac{2\lambda-1}{2p^2}q_{2\lambda-2,j}
\end{equation}
for the needed integral. Using this relation one can fore-bode the final result
presented in Eq.(\ref{q2lambda}) and prove it easily by induction to $\lambda$. 
%
%
\section{Special sums for the boundary behavior}
\label{appendixPojp0}
The $P_{\text{o}}(j,p,0)$ quantity is the coefficient of the error
function resulted from the integral $q_{2\lambda,j}$, see
Eq.(\ref{}). By this observation one can do the calculation on the
following way:
\begin{eqnarray}&&
\sum_{j=1}^dP_{\text{o}}(j,p,0)=\nonumber\\&&
\sum_{j=1}^d\sum_{\lambda=0}^{\lfloor\frac{\ell_{\text{max}}}{2}\rfloor}
\sum_{k=0}^{\lfloor\frac{\ell_{\text{max}}-1}{2}\rfloor}
C(j,2\lambda+2k,k)f(\lambda)=\nonumber\\&&
\sum_j^dC(j,0,0)+
\sum_{j=1}^d
\sum_{t=1}^{\left[\frac{\ell_{\text{max}}}{2}\right]}
\sum_{\lambda=0}^t
C(j,2t,t-\lambda)\frac{(2\lambda-1)!!}{2^{\lambda}}\nonumber\\&&=
\sum_{j=1}^d\frac{2A_{j(00)}}{p^2\mathcal{U}}+\nonumber\\&&
\sum_{j=1}^d
\sum_{t=1}^{\left[\frac{\ell_{\text{max}}}{2}\right]}
(-1)^tp^{-3}A_{j(2t0)}\frac{2\pi}{2p^2\mathcal{U}}\frac{\mathcal{N}_{2t}}{p^{2t}}\sqrt{\frac{4t+1}{4\pi}}
\sum_{\lambda=0}^tS(t,\lambda)\nonumber\\&&
\label{lastterm}
\end{eqnarray}
where
\begin{eqnarray}&&
S(t,\lambda)=f(\lambda)\tilde{\alpha}_{t-\lambda}(2t) (t-\lambda)!(-1)^{\lambda},\nonumber\\[1mm]&&
\tilde{\alpha}_{t-\lambda}(2t)=\frac{\prod_{i=0}^{t-\lambda}(t+1-i)(t+\frac{1}{2}-i)}{(t+1)(t+\frac{1}{2})(t-\lambda)!(t-\lambda)!}.
\end{eqnarray}
The first term in Eq.(\ref{lastterm}) vanishes by the charge neutrality conditions stated in Eq.(\ref{ewaldneutrality}).
Using the above expression for $\tilde{\alpha}_{t-\lambda}(2t)$ one gets
\begin{equation}
\begin{array}{ll}
S(t,0)=\frac{(2t-1)!!}{2^t}&\lambda=0\\[2mm]
S(t,\lambda)=\frac{t!}{\lambda!}\frac{(2t-1)!!}{2^t(t-\lambda)!}(-1)^{\lambda}&\lambda>0
\end{array}
\end{equation}
and then one can easily establish that
\begin{equation}
\begin{array}{ll}
t\text{\ is odd:}& S(t,\lambda)=-S(t,t-\lambda)\quad \lambda=0,\ldots,\frac{t-1}{2}\\[2mm]
t\text{\ is even:}&S(t,\lambda)=S(t,t-\lambda)\quad \lambda=0,\ldots,\frac{t}{2}-1.
\end{array}
\end{equation}
By this property for odd $t$ it is trivial that the last summation in Eq.(\ref{lastterm}) gives zero. For even $t$
($t=2\delta$)
\begin{eqnarray}
\sum_{\lambda=0}^tS(2\delta,\lambda)&=&\frac{(4\delta-1)!!}{4^{\delta}}+\sum_{\lambda=1}^{2\delta}\frac{(2\delta)!}{\lambda!}
\frac{(4\delta-1)!!}{4^{\delta}(2\delta-\lambda)!}(-1)^{\lambda}\nonumber\\
&=&\frac{(4\delta-1)!!}{4^{\delta}}\sum_{\lambda=0}^{2\delta}\binom{2\delta}{\lambda}(-1)^{\lambda}=0.
\end{eqnarray}

Checking the boundary behavior of the Ewald potential in
Eq.(\ref{asymptotics1}) one needs  the equation
\begin{equation}
\sum_{t=0}^{\left[\frac{\ell_{\text{max}}-1}{2}\right]}\sum_{\lambda=0}^t
f(\lambda+1)C(j,2t+1,t-\lambda)=C(j,1,0)f(1),
\label{asymptotics2}
\end{equation}
which can be proven on a similar way given above using the
\begin{equation}
\tilde{\alpha}_{t-\lambda}(2t+1)=\frac{\prod_{i=0}^{t-\lambda}
(t+\frac{3}{2}-i)(t+1-i)}{(t+\frac{3}{2})(t+1)(t-\lambda)!(t-\lambda)!}
\end{equation}
expression for the expansion coefficients of the hypergeometric function
in Eq.(\ref{sphericalbyhypergeometric2}).
%
%
\section{The one dimensional periodic case}
\label{1dperiodicity}
This appendix shortly sketches how the method inherited from the above presented treatment
apply to the one dimensional periodic case. The general
solution, characterized by the multipole moments of the unitcell and based on
the Poisson summation formula, is already presented in the literature
see Refs.\cite{Langridge,Delhalle}. The Ewald method for point-like Coulomb and
dipole-dipole interactions was worked out first by Porto\cite{Porto}.

Fourier expansions in the Poisson equation yields the 2D modified Helmholtz equation
\begin{equation}
\left(\frac{\partial^2}{\partial x^2}+\frac{\partial^2}{\partial y^2}
-K^2\right)\phi^{\text{Ewald}}(x,y,K)=\rho^{\text{Ewald}}(x,y,K).
\label{2DHelmholtz}
\end{equation}
The equation requires different treatments for $K\ne0$ and for the case
of $K=0$. The $K\ne0$ case is solved with the help of the Green function
\begin{equation}
G(K,\BM{\rho}-\BM{\rho}')=-\frac{1}{2\pi}K_0(K|\BM{\rho}-\BM{\rho}'|),
\end{equation}
where $K_0$ is the modified Bessel function of the second kind ($K_n$)
for $n=0$. Unlike in the 2D periodic case now one can use the Laplace
transform of the Green function
\begin{eqnarray}&&
\phi^{\text{Ewald}}(x,y,K)=\left(-\frac{1}{L2\pi}\right)\times\nonumber\\&&
\int_{\mathbbm{R}^2}dx'dy'\left[\int_{0^+}^{\infty}dt\frac{e^{-K^2/(4t)}}{2t}
e^{-t(x-x')^2-t(y-y')^2}\right]\times\nonumber\\&&
\sum_{j=1}^d\int_{\mathbbm{R}}dz'\ \rho_j^{\text{Ewald}}(\mathbf{r}'-\mathbf{s}_j)e^{-iKz'}
\end{eqnarray}
and then apply the Fourier transforms ($t>0$ by the Laplace transform)
\begin{equation}
e^{-t(\tau-\tau')^2}=\frac{1}{\sqrt{2\pi}}\int_{\mathbbm{R}}d\omega\ 
\frac{e^{-\omega^2/(4t)}}{\sqrt{2t}}e^{i\omega(\tau-\tau')}
\end{equation}
for the two exponential terms in the bracket. 
Of course, the $dt$ integration
behind the $\omega_x$ and $\omega_y$ integrals can also be executed and
results the two dimensional version of Eq.(\ref{mainequation2}), but its
calculation turns to difficulties after applying the already introduced
multipole shaped Ewald densities. Executing the integrations in a
different order
\begin{equation}
\int_0^{\infty} dt\ \int_{\mathbf{R}} d\omega_x\ \int_{\mathbf{R}} d\omega_y\ \int_{\mathbf{R}^3}d^3r'
\end{equation}
allows one to introduce the Fourier vector
$\mathbf{K}^{\text{T}}(\omega_x,\omega_y)=(\omega_x,\omega_y,K)$ similar
to that is used above in Eq.(\ref{newKvector2D}). The cylindrical
arrangement of the system suggests the application of Eq.(\ref{sphericalbyhypergeometric2})
to represent the spherical harmonics in the calculations. The phase
factor in $Y_{\ell,m}$ can be given as
\begin{equation}
e^{im\varphi}=\frac{(\omega_x+i\text{sgn}(m)\omega_y)^{|m|}}{\rho^{|m|}},
\end{equation}
which leads after its binomial expansion to the general result of
the $d\omega_xd\omega_y$ integral
\begin{eqnarray}
\mathcal{I}_{\alpha,\beta}(t)&=&
(i)^{\alpha\pm2\text{mod}(\alpha,2)}\left(\frac{\partial^{\alpha}}{\partial(x-s_{jx})^{\alpha}}\mathcal{I}_x(t)\right)\times\nonumber\\&&
(i)^{\beta\pm2\text{mod}(\beta,2)}\left(\frac{\partial^{\beta}}{\partial(y-s_{jy})^{\beta}}\mathcal{I}_y(t)\right),
\label{omegaxomegay_integral}
\end{eqnarray}
where
\begin{equation}
\mathcal{I}_i(t)=\frac{\sqrt{\pi}}{\sqrt{\frac{1}{p^2}+\frac{1}{t}}}\exp{\left(-\frac{p^2t(i-s_{ji})}{p^2+t}\right)}
\quad i\in\{x,y\}
\end{equation}
with its special case
\begin{equation}
\mathcal{I}_{0,0}(t)=\mathcal{I}_x(t)\mathcal{I}_y(t).
\end{equation}
Having these integrals one should turn to the $dt$ integration, which
shows the following form
\begin{equation}
\int_0^{\infty}dt\ \frac{\exp{\left(-\frac{K^2}{4t}-\frac{K^2}{4p^2}\right)}}{4t^2}\mathcal{I}_{\alpha,\beta}(t).
\end{equation}
In general the $dt$ integration leads to the so called leaky aquifer
function $W_n$.
Especially for the case of $L=(0,0)$ - only monopole compensation -
where $\alpha=\beta=0$ the zeroth order function $W_0$ is obtained
\begin{eqnarray}&&
\int_0^{\infty}dt\
\frac{\exp{\left(-\frac{K^2}{4t}-\frac{K^2}{4p^2}\right)}}{4t^2}\frac{\pi
  p^2t}{t+p^2}\times\nonumber\\&&
\exp{\left(-\frac{t}{p^2+t}\left[p^2(x-s_{jx})^2+p^2(y-s_{jy})^2\right]\right)}=\nonumber\\&&
\frac{\pi}{4}W_0\left(K^2/(4p^2),p^2(x-s_{jx})^2+p^2(y-s_{jy})^2\right).
\end{eqnarray}
Appropriate numerical representation of the leaky aquifer functions is discussed by
the F.E. Harris\cite{Harris-W-1,Harris-W-2} and
E.S. Kryachoko\cite{Kryachko}.

To obtain the solution of the equation with $K=0$ one can follow
the method of separation of variables after some art of manipulations.
The equation is rewritten in cylindrical coordinate system
and the solution, according to the assumption of the finite number of
non-zero electrostatic multipole moments, is looked for in the form of
\begin{equation}
\phi^{\text{Ewald}}(x,y,0)
=\sum_{j=1}^d\sum_{\ell,m}^{\ell_{\text{max}}}A_{jL}f_{L}(\varrho_j)
\zeta_{m0}e^{im\varphi_j},
\end{equation}
where $\zeta_{m0}$ is already introduced by
Eq.(\ref{sphericalbyhypergeometric2}) and
$\varrho_j=\sqrt{(x-s_{jx})^2+(y-s_{jy})^2}$.
Applying again the appropriate
(Eq.(\ref{sphericalbyhypergeometric2})) representation of the $Y_L$
functions, the original equation yields a second order linear
inhomogeneous ordinary differential equation for $f_{jL}$,
\begin{eqnarray}&&
f''_L+\frac{1}{\varrho}f'_L-\frac{m^2}{\varrho^2}f_L=\frac{\mathcal{N}_{\ell}}{L}\sqrt{\frac{2\ell+1}{4\pi}\frac{(\ell+|m|)!}{(\ell-|m|)!}}
\frac{1}{2^{|m|}|m|!}\times\nonumber\\&&
\sum_{k=0}^{\lfloor\frac{\ell-|m|}{2}\rfloor}\tilde{\alpha}_k(\ell,m)(-1)^ke^{-\varrho^2p^2}\varrho^{2k+|m|}\frac{\Gamma(\frac{2\lambda+1}{2})}{p^{2\lambda+1}},
\end{eqnarray}
where $2\lambda=\ell-2k-|m|$. Only the $\text{mod}(\ell-|m|,2)=0$ case
has nonzero contribution. The general solutions of the
homogeneous equation are
\begin{equation}
\begin{array}{rll}
f_{\ell,0}(\varrho)=&\mathcal{A}_0\displaystyle\log{(p\varrho)}+\mathcal{B}_0,&m=0\\[3mm]
f_{\ell,|m|}(\varrho)=&\mathcal{A}_m(p\varrho)^{-|m|}+\mathcal{B}_m(p\varrho)^{|m|},&|m|>0
\end{array}
\end{equation}
The solutions of the inhomogeneous equation for both $m=0$ and
$|m|>0$ can be obtained by the constant variation method. For example,
for $m=0$
\begin{equation}
\begin{array}{rl}
\mathcal{A}_0(\varrho)=&-\displaystyle\frac{1}{L\sqrt{\pi}}e^{-p^2\varrho^2},\\[6mm]
\mathcal{B}_0(\varrho)=&\displaystyle\frac{1}{L\sqrt{\pi}}\left(-\frac{\text{Ei}(-p^2\varrho^2)}{2}+e^{-p^2\varrho^2}\log{(p\varrho)}\right)
\end{array}
\end{equation}
which results
\begin{eqnarray}
\phi^{\text{Ewlad}}(x,y,K)&=&
\sum_{j=1}^d\frac{q_j}{4\pi L}\Big[\Gamma(0,p^2\varrho_j^2)-\nonumber\\&&
\frac{\mathcal{A}_0}{\sqrt{4\pi}}\log{(p^2\varrho_j^2)}\Big]
\end{eqnarray}
by the charge neutrality.
Because of the identity (see Gradshteyn\cite{Gradshteyn} 8.212/1.)
\begin{eqnarray}&&
\Gamma(0,x)-\frac{\mathcal{A}_0}{\sqrt{4\pi}}\log{(x)}=
-\log{(x)}\left(1+\frac{\mathcal{A}_0}{\sqrt{4\pi}}\right)-\nonumber\\&&
\gamma-\int_0^xdt\ \frac{e^{-t}-1}{t},\quad\quad x>0,
\end{eqnarray}
the choice of $\mathcal{A}_0=-\sqrt{4\pi}$ together with the charge neutrality
ensures the required vanishing
potential at the infinity (open boundary).

As it is shown above in the simple example the obtained results are
the same what are already presented by Langridge and co-workers\cite{Langridge}
or by Porto\cite{Porto} but of course, the followed path is methodologically different.

%
%
\begin{acknowledgements}
This work was supported by the European Commission (RTN $\psi_k$ $f$-electron). 
The helpful discussions with Klaus Koepernik (IFW Dresden) and Manuel
Richter (IFW Dresden) are gladly acknowledged.
\end{acknowledgements}
%
%
%
\bibliography{ewald}
\bibliographystyle{apsrev}
%
%
\end{document}